\documentclass[nofootinbib,aps,10pt,twocolumn]{revtex4-1}

\usepackage{amsmath,amssymb,bm,epsfig}
\usepackage{color}
\usepackage{hyperref} 
\usepackage{graphicx}
\usepackage{xcolor,colortbl}
\RequirePackage{lineno}
\definecolor{Gray}{gray}{0.85}
\newcolumntype{a}{>{\columncolor{Gray}}c}

\def \beq{\begin{equation}}
\def \eeq{\end{equation}}
\def \beqa{\begin{eqnarray}}
\def \eeqa{\end{eqnarray}}

\begin{document}

\title{Interplay of longitudinal and transverse expansion  in the kinetic  dynamics of heavy-ion collisions }

\author{Piotr Bo{\.z}ek}
\email{piotr.bozek@fis.agh.edu.pl}
\affiliation{AGH University of Science and Technology,\\ 
Faculty of Physics and Applied Computer Science,\\
aleja Mickiewicza 30, 30-059 Krakow, Poland}

\begin{abstract}

  The early dynamics in heavy-ion collisions  involves a rapid,
  far from equilibrium evolution.   This early pre-equilibrium stage of the
  dynamics can be modeled using kinetic equations. The effect of this
  pre-equilibrium stage on
  final observables derived from transverse momenta of emitted particles
  is small. The kinetic equations in the
  relaxation  time approximation for a non-boost invariant system are solved.
  The asymmetry of the flow with respect to the reaction plane at
  different rapidities is found to be 
  very sensitive to the degree of non-equilibrium in the evolution.
  This suggests that the rapidity odd directed flow could be studied
  to identify the occurrence of non-equilibrium effects and  to estimate the
  asymmetry of the pressure between
  the longitudinal and transverse directions in the collision.

\end{abstract}

\maketitle

\section{Introduction}

The experiments preformed at  the Relativistic Heavy Ion Collider and the
Large Hadron Collider aim at studying the properties of dense matter created in
heavy-ion collisions. The properties of the expanding matter can be extracted
by
analyzing the collective flow of matter. This
requires precise modeling of the collision dynamics
\cite{Ollitrault:2010tn,Teaney:2009qa,Heinz:2013th,Gale:2013da}, using the relativistic viscous hydrodynamics. This
rises the question of the range of applicability of the hydrodynamic approach,
especially at the early stages of the collision, where the system is expected
to evolve far from local thermal equilibrium.

The hydrodynamic model can be combined with a non-equilibrium early stage. The
simplest possibility is to use a free-streaming early evolution coupled with a
latter relativistic hydrodynamic expansion \cite{Broniowski:2008qk}.
A free-streaming  pre-equilibrium stage for the viscous  relativistic 
hydrodynamics has been used in detailed quantitative analyses of collective
observables \cite{Bernhard:2019bmu,JETSCAPE:2020shq,Nijs:2020ors}.
Early evolution going beyond the free-streaming dynamics has been implemented
using a linearized response approach for general initial conditions
\cite{Kurkela:2018vqr,Gale:2021emg,daSilva:2022xwu}, by a solution of
the kinetic equation in the relaxation time approximation
\cite{Ambrus:2022koq,Ambrus:2022qya,Liyanage:2022nua},
or a numerical relativity solution to AdS/CFT  \cite{vanderSchee:2013pia}.

The implementation of a short no-equilibrium stage in the dynamics allows
to extend the range of applicability of the model to early times, where
relativistic viscous hydrodynamics is expected to break down.
Combining an early
non-equilibrium  stage with hydrodynamics allows to make more precise quantitative prediction for collective observables. However, results of simulations show, that the  effect of the pre-equilibrium stage on final observables
is limited. The
corrections to the final observables for the collective flow in the transverse
plane,
when using free-streaming, kinetic
equations, or naive hydrodynamics in the early stage,
differ only at a few percent
level \cite{Kurkela:2018vqr,daSilva:2022xwu,Ambrus:2022koq}.  This is a
result of an universality in the transverse flow \cite{Vredevoogd:2008id}.

Different physical scenarios for the early stage dynamics, ranging from
free-streaming expansion to ideal fluid dynamics, involve very different
behavior of the local particle distribution, with very different asymmetry
between the longitudinal and transverse pressures. It would be interesting to
find collective flow variables more sensitive to the pressure asymmetry in
the dynamics than the flow observables in the transverse plane. It has been noticed that the rapidity dependent directed flow with respect to the reaction plane 
may be sensitive the early stage pressure asymmetry in hydrodynamics
\cite{Bozek:2010aj}.

In this paper a numerical solution of the kinetic equation in the
relaxation time approximation is found in a non-boost invariant geometry. In
order to study the sensitivity of the collective flow to the pressure asymmetry, one needs a model that has
\begin{itemize}
\item{}  a non-boost invariant dynamics, 
\item{}  full treatment of the longitudinal and transverse momenta of
  particles in the kinetic equation,
\item{} and a least one non-trivial transverse direction.
\end{itemize}
 I solve the kinetic equation for massless particles in the relaxation
 time approximation with one longitudinal and one transverse direction.
 The two directions define the reaction plane of the collision.
 The system is assumed to be homogeneous in the second transverse direction.
 The initial conditions used allow for the formation of a
 non-zero, rapidity dependent
 directed flow. I study the sensitivity of the transverse flow on the pressure asymmetry (the value of the relaxation time). The new result
 is the study of the dependence of 
 the rapidity 
 dependent asymmetry of the flow in the transverse direction on the equilibration rate.
 
\section{Kinetic equation}

The kinetic equation for the phase space distribution $f(t,\vec{x},\vec{p})$
in the relaxation time approximation has the form
\begin{equation}
  p^\mu\partial_\mu f = -\frac{u^\mu p_\mu}{\tau_R}\left( f-f_{iso}\right) \ ,
  \label{eq:rta}
\end{equation}
where for massless on-shell particles
$p^\mu=\left( p ,\vec{p} \right)$,  $p=|\vec{p}|$.
The  system relaxes to an isotropic distribution distribution $f_{iso}$
in the local rest frame. The isotropic  distribution $f_{iso}$ is not
necessarily the local equilibrium distribution and the approach is named
the isotropization time approximation \cite{Kurkela:2018ygx}. $u^\mu(t,\vec{x})$
is the  space-time dependent flow velocity of the local rest frame. The local rest frame is defined by the matching condition at each space-time position $(t,\vec{x})$
\begin{equation}
  T^{\mu\nu}u_\nu = \epsilon u^\mu \ ,
\end{equation}
where $\epsilon$ is the local energy density and the energy-momentum tensor is
\begin{equation}
  T^{\mu \nu} = \int \frac{d^3p}{(2\pi)^3\  p }p^\mu p^\nu f(t,\vec{x},\vec{p}) \ .
  \end{equation}

The solution of the kinetic equation can be simplified \cite{Kurkela:2018ygx} using the moment
\begin{equation}
  F(t,\vec{x},\Omega_p)=\int \frac{p^3 dp}{2 \pi^2} f(t,\vec{x},\vec{p})
  \label{eq:moment}
  \end{equation}
of the distribution function. The distribution function $F$ depends on the spherical angles  $\Omega_p=(\phi,\theta)$ of the particle  momentum vector $\vec{p}$.
It obeys the following kinetic equation
\begin{equation}
  \partial_t F + \vec{v} \partial_{\vec{x}} F = - \frac{u^\mu v_\mu}{\tau_R}\left( F-F_{iso}\right) 
\end{equation}
with \cite{Kurkela:2018ygx}
\begin{equation}
  F_{iso}(t,\vec{x}) = \frac{\epsilon(t,\vec{x})}{(u^\mu v_\mu)^4} \ .
  \label{eq:fiso}
\end{equation}
The particle velocity $v^\mu=(1,\vec{v})=(1,\vec{v}_T,v_z)=(1,\sqrt{1-v_z^2} \cos(\phi),\sqrt{1-v_z^2} \sin(\phi),v_z)$,  $v_z=\cos(\theta)$. The energy-momentum tensor can be reconstructed from the distribution function $F$
\begin{equation}
  T^{\mu\nu}= \int \frac{d\phi}{2\pi} \frac{d v_z}{2} v^{\mu} v^\nu F \ .
  \end{equation}

In Milne's coordinates, the proper time $\tau=\sqrt{t^2-z^2}$ and the space-time  rapidity  $\eta= \frac{1}{2}\ln\left(\frac{t+z}{t-z}\right)$, the kinetic equation takes the from
\begin{eqnarray}
& &  \left( \cosh \eta - v_z \sinh \eta\right) \partial_\tau F  + \frac{1}{\tau} \left(   v_z \cosh \eta - \sinh \eta \right) \partial_\eta F 
  \nonumber   \\
& +& \vec{v}_T \partial_{\vec{x}_T} F + \frac{u^\mu v_\mu}{\tau_R}\left(F-F_{iso} \right) = 0  \ . \label{eq:kin}
\end{eqnarray}
The model has been applied extensively in boost-invariant systems \cite{Kurkela:2018ygx,Kurkela:2018qeb,Kurkela:2019kip,Kurkela:2020wwb,Ambrus:2021fej,Liyanage:2022nua}. In that case, the kinetic equation at $z=0$  can be written as
\begin{eqnarray}
 & &  \partial_\tau F +\vec{v}_T\partial_{\vec{x}_T} F -\frac{1}{\tau}v_z (1-v_z^2)\partial_{v_z} F \nonumber \\
  &+ &\frac{4 v_z^2}{\tau} F + \frac{u^\mu v_\mu}{\tau_R}\left(F-F_{iso} \right) = 0 \ . \label{eq:k1d}
  \end{eqnarray}
Typically,  the evolution rapidly generates a strong asymmetry between the longitudinal and transverse width of the  momentum distribution. This motivates
the often used approximation
$F\propto \delta(v_z)$, which simplifies further the description.
Using this approach the kinetic response in the transverse plane can be studied.
In this paper I study non boost-invariant systems, so that the full dependence on the longitudinal coordinate must be kept. In that way the interplay between the effective longitudinal and transverse pressures in the dynamics can be studied. Therefore, the complete dependence of the distribution on $v_z$ is evolved in the numerical solution of Eq. \ref{eq:kin}.

\section{Solution in 1+1+1 dimensions}

In this paper I study the simplest case where a nontrivial  interplay
of the longitudinal and transverse expansions is possible. I assume that the distribution depends on the longitudinal and one transverse direction.
The distribution $F(\tau,x,\eta,\phi,v_z)$ is five-dimensional. Due to the symmetry of the system  it is sufficient to evolve only  angles  $0\le \phi \le \pi$. Alternatively, a moment expansion of the distribution
\begin{eqnarray}
 & &  F(\tau,x,\eta,\phi,v_z)= F_0(\tau,x,\eta,v_z)
   + 2 \cos(\phi) F_1(\tau,x,\eta,v_z)  \nonumber \\ & & + 2 \cos(2\phi) F_0(\tau,x,\eta,v_z)+ \dots 
  \end{eqnarray}
could be used.  In this paper the collision term is calculated iteratively,
so that the evolution equations (\ref{eq:kin}) for each value of $\phi$ separate and the numerical solution is more effective, when using directly the function
F with the full dependence in azimuthal angle.
Moreover, the moment expansion in a global reference frame could
introduce some dependence on the choice of the reference frame.

\begin{figure}
 \begin{center}
   \includegraphics[scale=0.5]{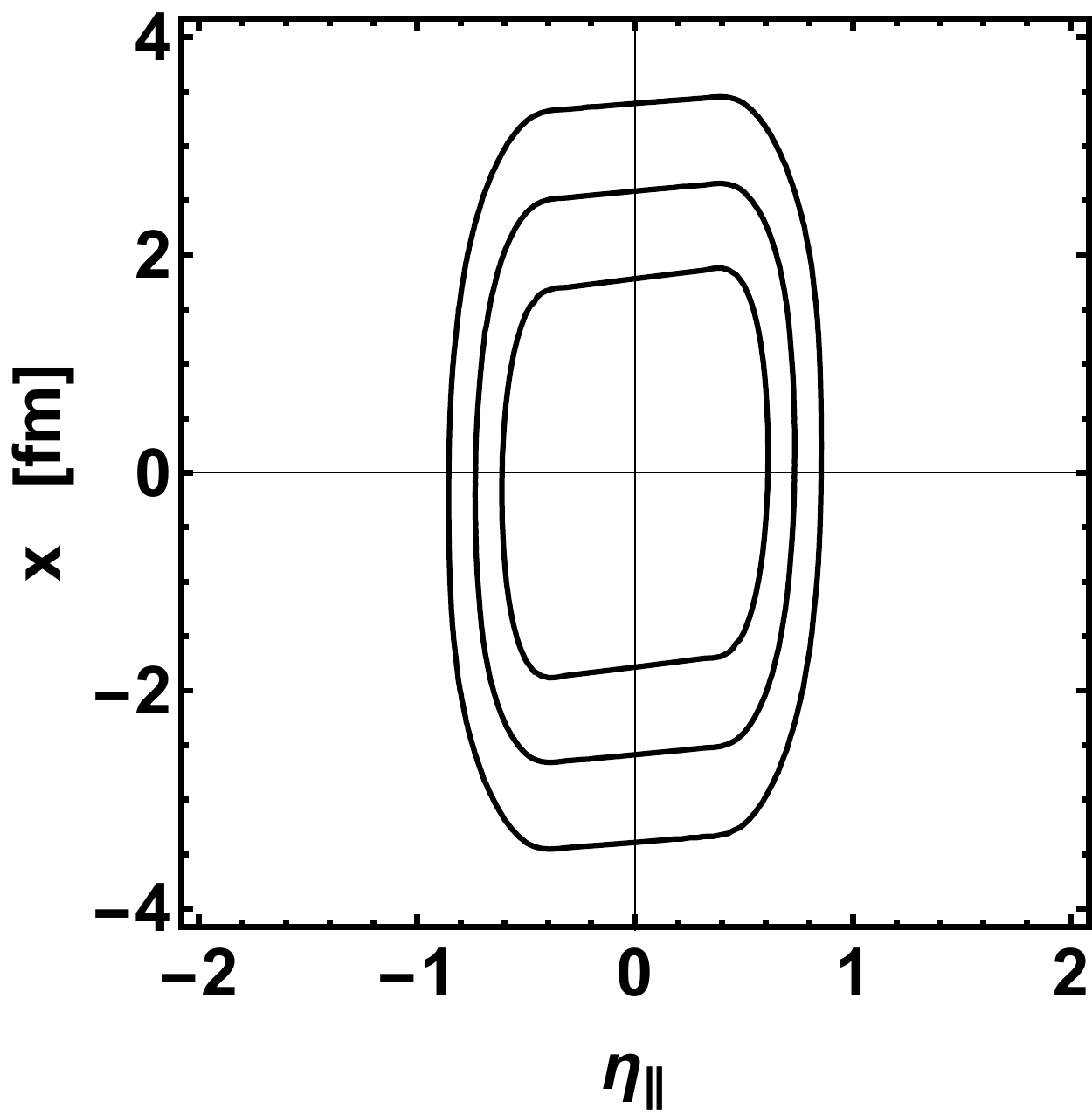}
   \vskip 0mm
   \caption{ Initial distribution of the energy density in the reaction plane,
     for Au-Au collisions, using the parametrization from \cite{Bozek:2022svy}.
   The three contours shown correspond to energy densities equal to $0.5$, $0.17$, and $0.017$ of the maximal energy density at the center of the fireball.}
 \label{fig:idens}
 \end{center}
\end{figure}

The initial conditions used in  this example correspond to  an initial
 fireball tilted away from the collision axis \cite{Bozek:2010bi}. In
the hydrodynamic model such initial conditions allow for the
generation of a rapidity dependent directed flow. The tilt of the initial density can originate from an asymmetric energy deposition in rapidity from participants from the target and from the projectile. The details and parameters of the model of the entropy distribution corresponding to Au-Au collisions can be found in
\cite{Bozek:2022svy}. I take the density for the impact parameter $9.6$~fm.
The contour plot of the resulting
initial energy density is show in Fig. \ref{fig:idens}. 
This initial energy density in the local rest frame  $\epsilon(x,\eta)$ is used to define the initial
distribution $F$. In order to model the case when in
the  local rest frame the angular distribution is
non-isotropic, I use a parametric distribution in the polar angle
\begin{equation}
  g(v_z)=N \exp\left( - \frac{v_z^2}{2 w_v^2}\right) \ ,
  \label{eq:inasy}
\end{equation}  
with $w_z=0.3$ and $\int_{-1}^{1} g(v_z) dv_z =1$. This corresponds to a
pressure asymmetry between the longitudinal and transverse pressure $P_L/P_T \simeq 0.2$. Simulations show that the dynamics depends very weakly on the precise form of the initial
asymmetric distribution $g(v_z)$  in the longitudinal direction, as long as it leads to the same value of the ratio  $P_L/P_T$. The form of the distribution $g(v_z)$ in a moving frame is obtained by boosting the phase-space distribution in Eq. \ref{eq:moment}.
The resulting $g(v_z)$ distribution  boosted in the longitudinal direction by rapidity $Y$ is
\begin{equation}
g(v_z,Y)= \frac{g\left( \frac{v_z \cosh(Y)-\sinh(Y)}{\cosh(Y)-v_z\sinh(Y)} \right)}{\left(\cosh(Y)-v_z\sinh(Y) \right)^4}  \ . \label{eq:iniall}
\end{equation}
In the initial state I assume the presence of a Bjorken flow $Y=\eta$ and no
transverse flow $u^\mu=\left( \cosh(Y),0,0,\sinh(Y) \right)$. The  distribution at the initial time $\tau_0$ is
\begin{equation}
  F(\tau_0,x,\eta,\phi,v_z)= \epsilon(\tau_0,\eta,x)  g(v_z,\eta)  \ .
\end{equation}
Obviously, for isotropic initial conditions ($P_L=P_T$), one  puts $g(v_z)=1$ and Eq. (\ref{eq:fiso}) is recovered. 

\begin{figure}
 \begin{center}
   \includegraphics[scale=0.5]{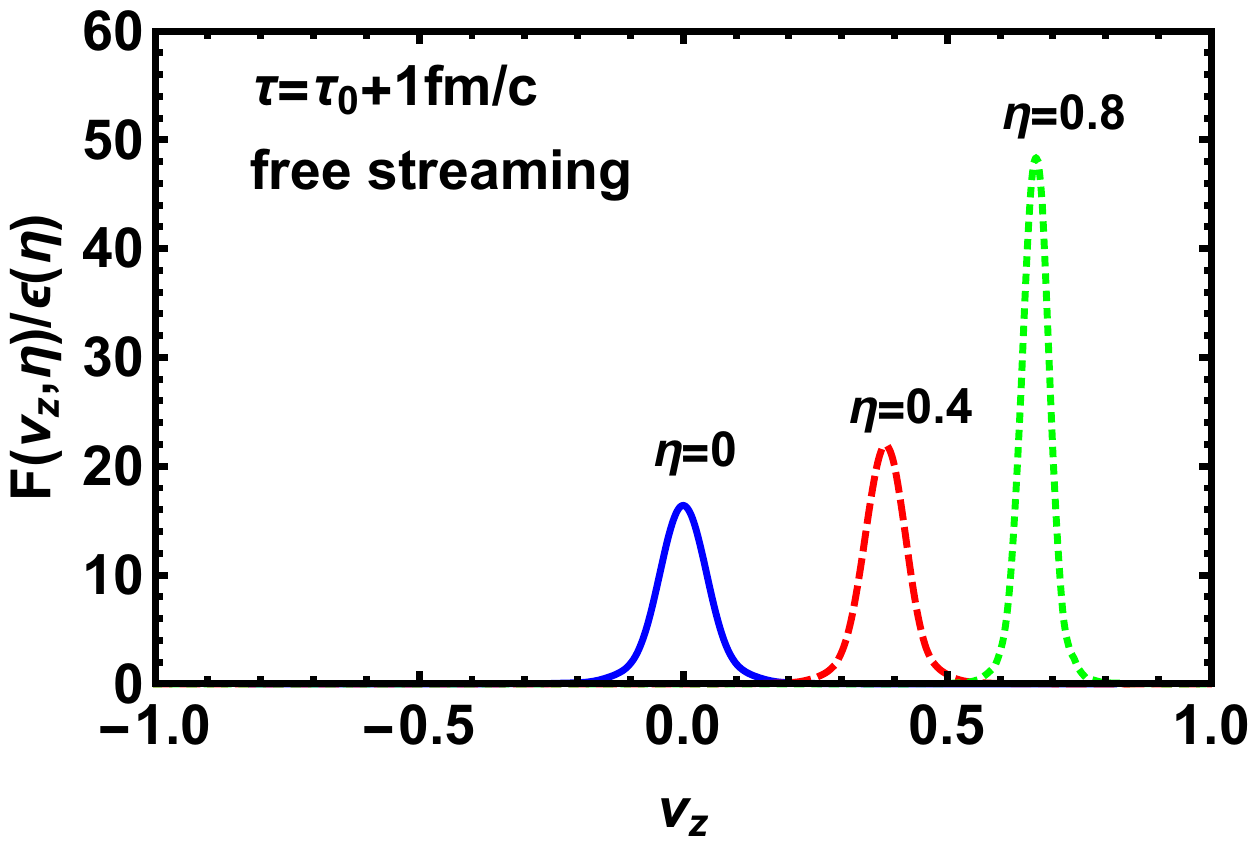}
   \vskip 0mm
 \caption{ The free-streaming   distribution  $F$ averaged over the azimuthal angle $\phi$ at $\tau=\tau_0+1$fm/c and $x=0$, at three different space-time rapidities, for the asymmetric initial distribution (\ref{eq:inasy}). }
 \label{fig:isoboost}
 \end{center}
\end{figure}

A numerical evolution of the distribution $F$ in the global reference frame
is computationally  demanding. The distribution $F$ in momentum variables $\phi$ and $v_z$ remains smooth in the local rest frame. After a boost to the frame moving with
large velocity the distributions get sharply peaked. An example is shown in Fig. \ref{fig:isoboost}. The  distribution, which is smooth
in the local rest frame ($\eta=0$),
is strongly varying  in the global reference frame.
Such distribution require the use of very fine interpolations grids, especially at large rapidities.

\begin{figure}
 \begin{center}
   \includegraphics[scale=0.48]{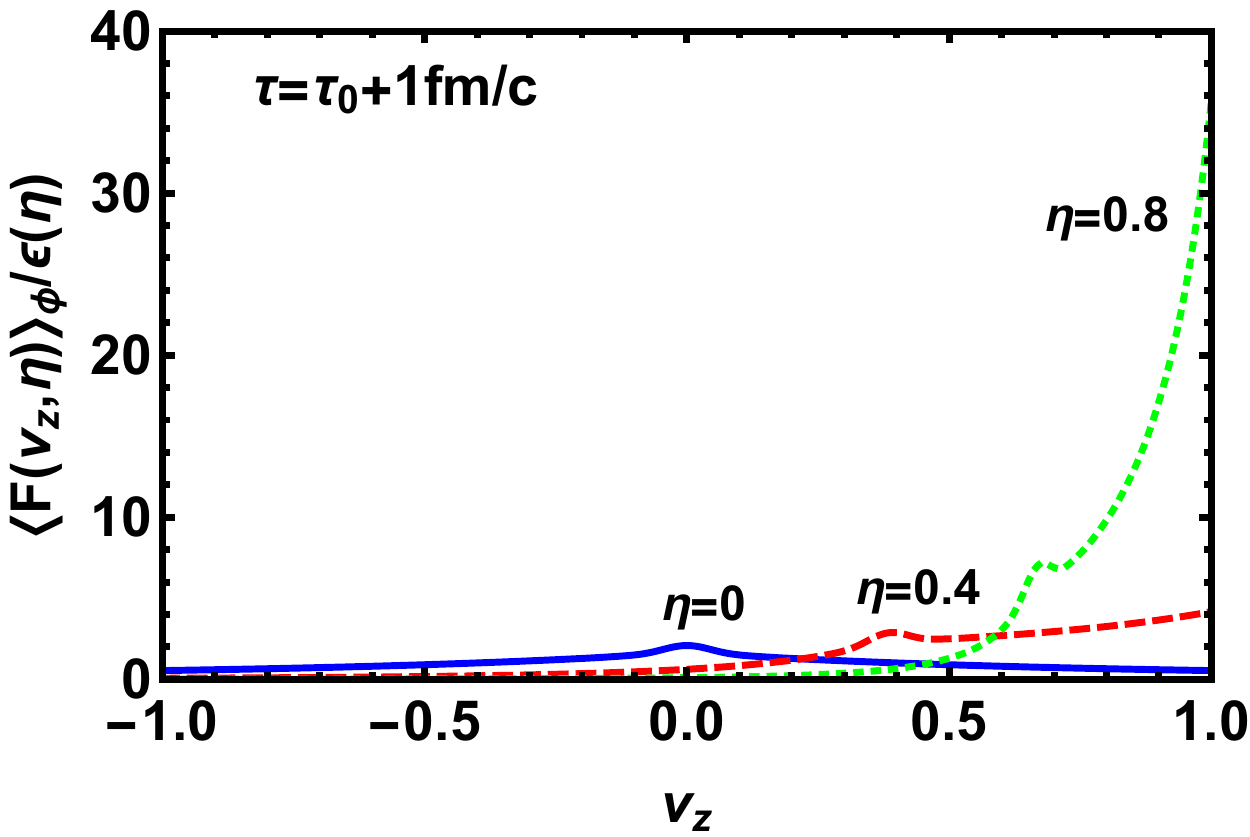}
   \vskip 0mm
   \caption{ The  distribution  $F$ averaged over the azimuthal angle $\phi$ at $\tau=\tau_0+1$fm/c and $x=0$, at three different space-time rapidities, for the asymmetric initial distribution (\ref{eq:inasy}).  The relaxation time
     corresponds $\eta/s=0.08$. }
 \label{fig:frelax}
 \end{center}
\end{figure}

\begin{figure}
 \begin{center}
   \includegraphics[scale=0.5]{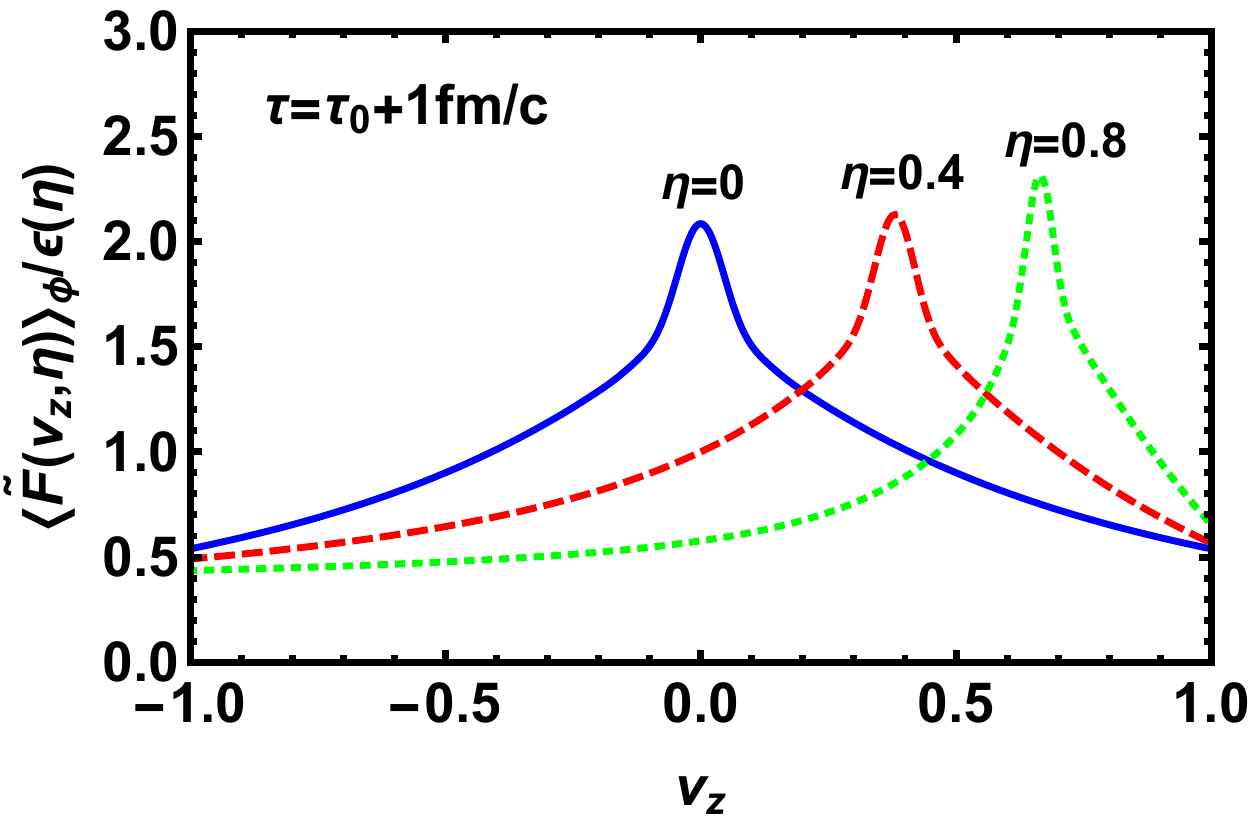}
   \vskip 0mm
 \caption{ Same as in Fig. \ref{fig:frelax} but for the rescaled distribution $\tilde{F}$. }
 \label{fig:ftilderelax}
 \end{center}
\end{figure}

In the case, when the dominant flow is the Bjorken flow in the longitudinal direction, it is advantageous to use a rescaled distribution $\tilde{F}$ distribution defined as
\begin{equation}
  F(\tau,x,\eta,\phi,v_z)=  \frac{ \tilde{F}(\tau,x,\eta,\phi,v_z) }{\left(\cosh(\eta)-v_z\sinh(\eta) \right)^4} \ ,
\end{equation}
The kinetic evolution equation for $\tilde{F}$ is
\begin{eqnarray}
  & &  \left( \cosh \eta - v_z \sinh \eta\right) \partial_\tau \tilde{F}  + \frac{1}{\tau} \left(  v_z \cosh \eta- \sinh \eta  \right) \partial_\eta \tilde{F} 
  \nonumber   \\
  & + & 4 \frac{ \left(  v_z \cosh \eta- \sinh \eta\right)^2}{\left( \cosh \eta - v_z \sinh \eta \right) \tau} \tilde{F} + \vec{v}_T \partial_{\vec{x}_T} F  \nonumber \\
  &+ & \frac{u^\mu v_\mu}{\tau_R}\left(\tilde{F}- \epsilon(\tau,\eta,x) \frac{\left( \cosh \eta - v_z \sinh \eta\right)^4}{ \left(  u^\mu v_\mu\right)^4 } \right) = 0  \ . \label{eq:kintilde}
\end{eqnarray}
In Fig. \ref{fig:frelax} is shown the distribution function $\langle F \rangle_\phi=\int\frac{d\phi}{2\pi}F(\tau,x,\eta,\phi,v_Z)$. An accurate numerical evolution of that function requires a fine numerical grid. On the other hand, the rescaled distribution $\langle \tilde{F}\rangle _\phi$ (Fig. \ref{fig:ftilderelax}) can be easily discretized. 

The evolution Eq. \ref{eq:kintilde} separates for every azimuthal angle,
except for the flow $u^\mu$ and the rest frame energy density $\epsilon$.
The collision term is calculated iteratively, outside of the differential evolution equation. I have checked that, in the case of explicitly boost invariant and $x$ independent initial conditions,  the numerical solution  of Eq. \ref{eq:kintilde} is the same as obtained from  Eq. \ref{eq:k1d}. In this paper
I study the solution in  the space-time rapidity range $-0.8<\eta<0.8$
and for $\tau$ from $\tau_0=0.2$fm/c to $1.2$fm/c.

The kinetic equations for massless particles are independent on the initial
energy scale \cite{Kurkela:2018ygx}. The scale in the problem
is introduced via the relaxation time $\tau_R$. The relaxation time is assumed to depend on the energy density $\tau_R \propto \epsilon^{-1/4} \propto T^{-1}$, which corresponds to a constant value of the shear viscosity over
entropy ratio $ 5 \eta/s = \tau_R T$  \cite{Florkowski:2013lya}. I compare three calculations, the free-streaming evolution ($\tau_R=\infty$) and two evolutions with a collision term, with   relaxation times of the form 
$\tau_R=  \tau_C \left( \frac{\epsilon(\tau_0,0,0)}{\epsilon(\tau,x,\eta)}\right)^{-1/4}$
and $\tau_C=0.2$ or $0.4$ fm/c. The corresponding average relaxation times  (weighted with $\epsilon$) change from 
$0.22\ (0.44)$~fm/c at $\tau_0$ to $0.4\ (0.8)$~fm/c at $\tau=1.2$~fm/c. When
setting the scale by choosing the initial temperature at the center of the fireball to   $400$~MeV, $\tau_C=0.2$~fm/c and $0.4$~fm/c would corresponds to
$\eta/s=0.08$ and $0.16$ respectively. In the following, these two calculations are labeled
according to those shear viscosity to entropy  ratios $\eta/s=0.8$ and $0.16$.
One should remember that it is only a way to label the interaction strength and does not imply that viscous hydrodynamics apply.
The final  evolution time $\tau=1.2$~fm/c corresponds to $\tau/\tau_R \simeq 2$ - $4$  for simulations with $\tau_c=0.4$ and $0.2$ fm/c. 

\section{Results}

\begin{figure}
 \begin{center}
   \includegraphics[scale=0.5]{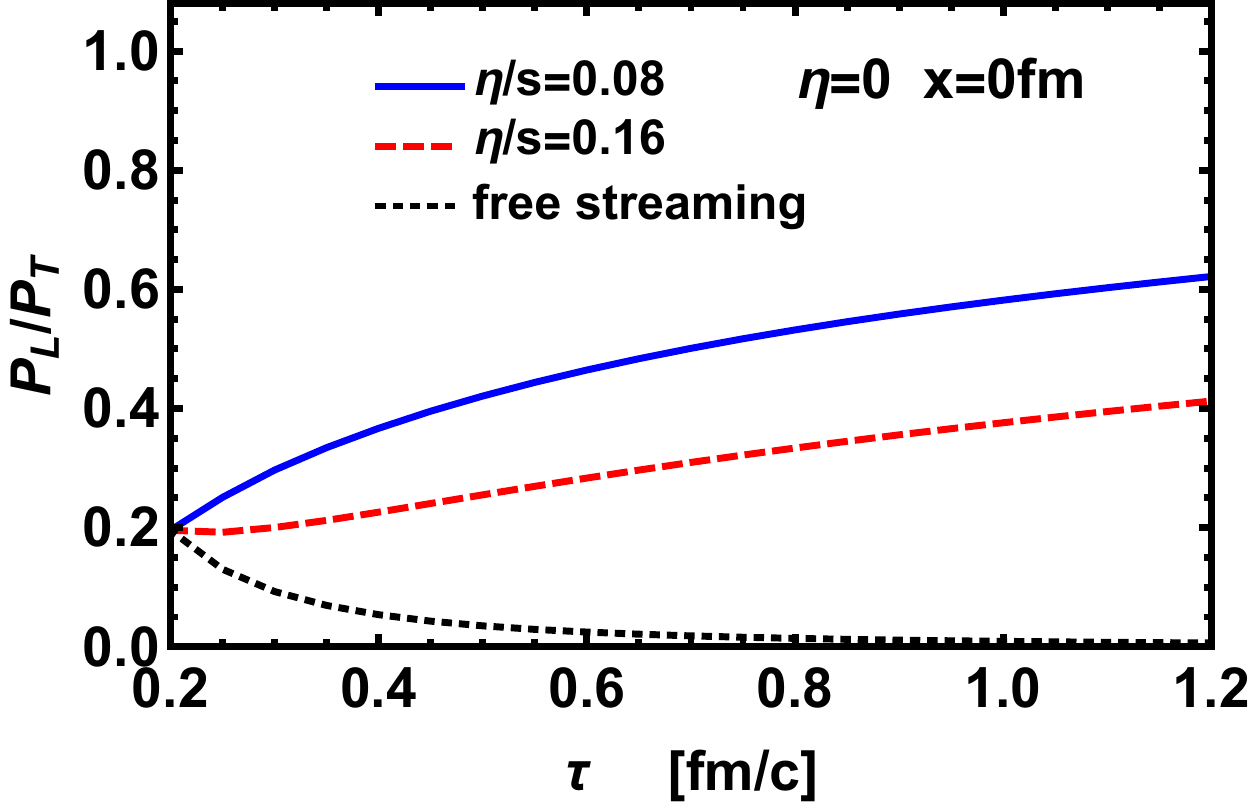}
   \vskip 0mm
 \caption{ Evolution of the ratio of longitudinal to transverse pressure at the center of the fireball ($\eta=0$, $x=0$). The solid, dashed and dotted lines represent the results of the kinetic evolution for $\eta/s=0.08$, $\eta/s=0.16$, and  the free-streaming evolution, respectively.}
 \label{fig:plptcenter}
 \end{center}
\end{figure}

\begin{figure}
 \begin{center}
   \includegraphics[scale=0.5]{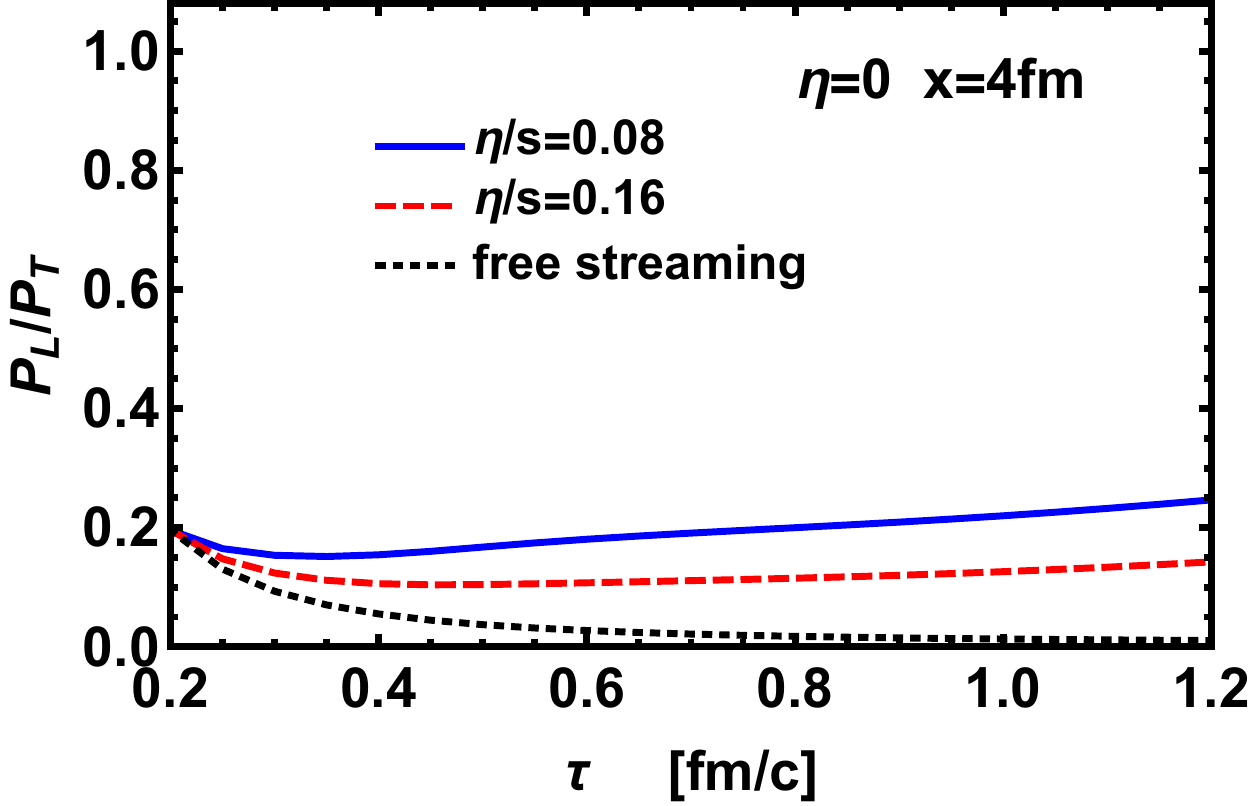}
   \vskip 0mm
 \caption{ Same as in Fig. \ref{fig:plptcenter} but for $\eta=0$ and $x=4$~fm/c. }
 \label{fig:plptx4}
 \end{center}
\end{figure}

The kinetic equations are evolved starting from the initial conditions (\ref{eq:iniall}), corresponding to an initial pressure asymmetry $P_L/P_T=0.2$.
The evolution of the pressure asymmetry
\begin{equation}
  \frac{P_L}{P_T}=\frac{2 \tilde{T}_{zz}}{\tilde{T}_{xx}+\tilde{T}_{yy}} \ ,
  \end{equation}
with $\tilde{T}$ being the energy momentum tensor in the local rest frame, is
shown in Figs. \ref{fig:plptcenter} and \ref{fig:plptx4}.
The three choices for the relaxation time lead to very different pressure
asymmetries. The free-streaming evolution results in a rapid increase of the pressure asymmetry. At the center of the fireball ($\eta=0$, $x=0$), the evolution with fast equilibration ($\eta/s=0.08$) results in an approach to pressure isotropy, with $P_L/P_T$ reaching $0.6$. For the collision rate corresponding to $\eta/s=0.16$ the pressure asymmetry remains significant, with $P_L/P_T\simeq 0.2 - 0.4$.

Generally, one expects that the relaxation time is increasing with decreasing energy density. In this calculation this is encoded in the formula
$\tau_R T = 5 \eta/s$ , with constant $\eta/s$. The relaxation time is larger at the edge of the fireball and non-equilibrium effects are more pronounced. The pressure asymmetry at the edge of the fireball remains large through the evolution of the system, also for the small viscosity case (Fig. \ref{fig:plptx4}).

\begin{figure}
 \begin{center}
   \includegraphics[scale=0.5]{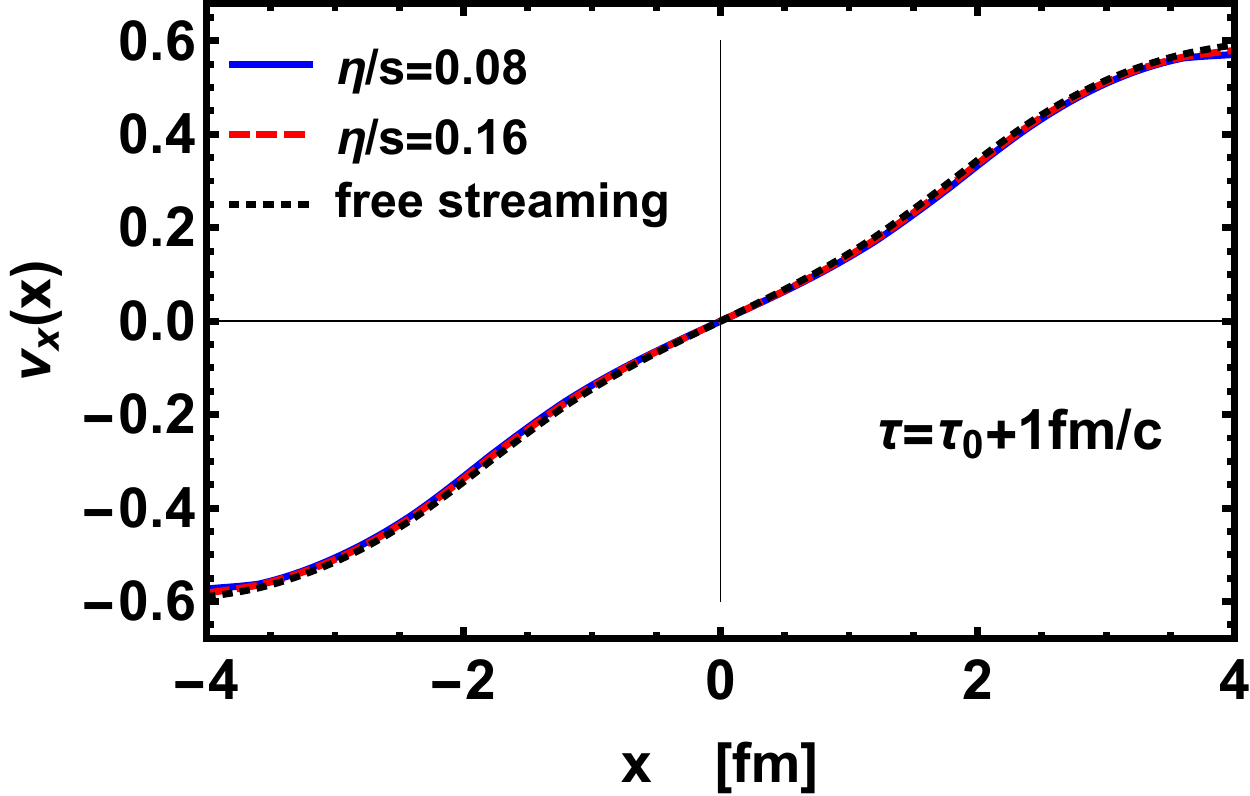}
   \vskip 0mm
 \caption{ The transverse velocity profiles at $\eta=0$ at $\tau=1.2$fmc/c (same symbols as in Fig. \ref{fig:plptcenter}).  }
 \label{fig:vxprofile}
 \end{center}
\end{figure}

In the chosen geometry, the transverse expansion of matter builds up the velocity $v_x=u_x/\gamma_T$, $\gamma_T=\sqrt{1+u_x^2}$. The transverse flow velocity profiles remains almost indistinguishable during the early evolution (Fig. \ref{fig:vxprofile}) for the three calculations.
This effect reflects the universality of the early transverse flow \cite{Vredevoogd:2008id}. The average transverse flow is $\langle v_T \rangle = \langle |v_x| \rangle$, with the average defined as
\begin{equation}
  \langle \dots \rangle =\frac{\int dx \dots \gamma_T(\tau,\eta,x) \epsilon(\tau,\eta,x) }{\int dx  \gamma_T(\tau,\eta,x) \epsilon(\tau,\eta,x)} \ . \label{eq:average}
\end{equation}

\begin{figure}
 \begin{center}
   \includegraphics[scale=0.5]{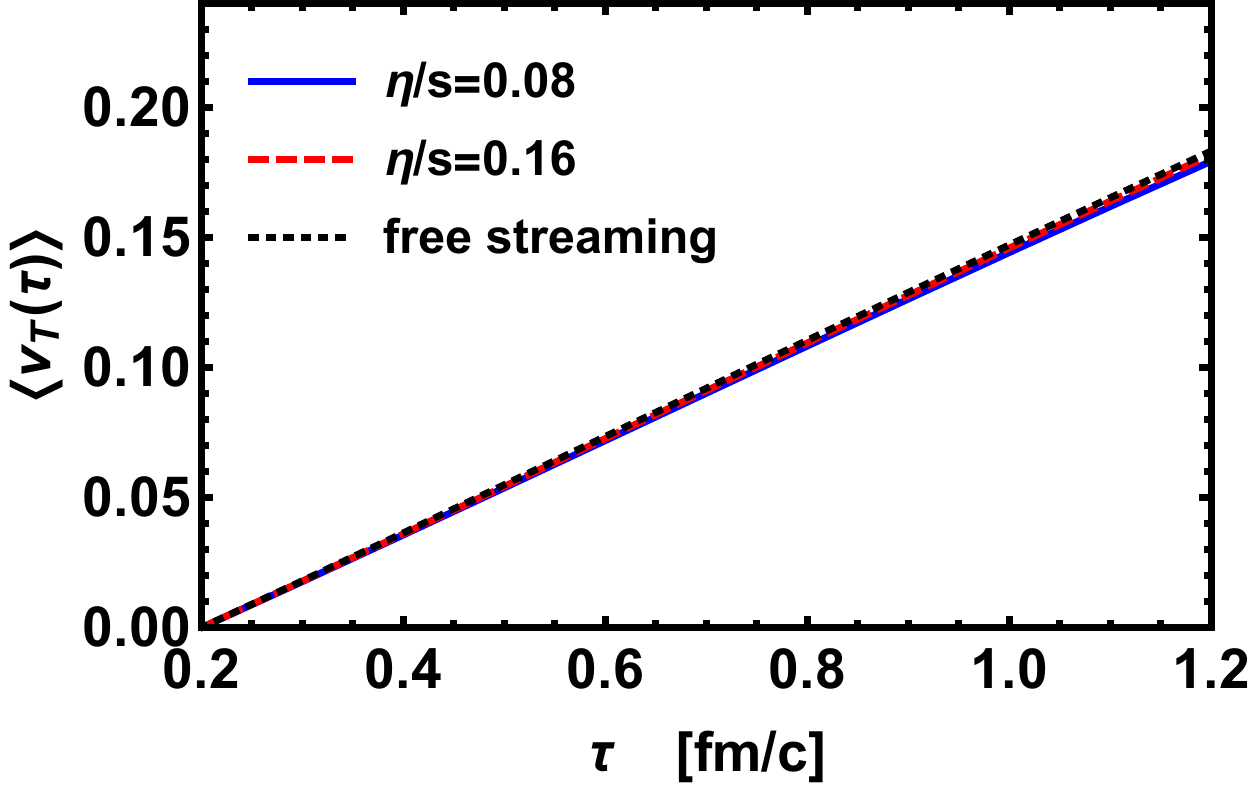}
   \vskip 0mm
 \caption{ The average transverse velocity at $\eta=0$ as a function of the evolution time  (same symbols as in Fig. \ref{fig:plptcenter}).  }
 \label{fig:vt}
 \end{center}
\end{figure}

The average transverse flow increases linearly with the evolution time $\tau-\tau_0$ (Fig. \ref{fig:vt}). Through the whole evolution the accumulated
average transverse flow is almost the same in the three calculations,
independently on the value of the relaxation time. The geometry in this study does not allow to calculate the elliptic flow coefficient.
Other studies have shown, that for small evolution times the kinetic evolution
with different realistic relaxation times,  or even  using naive hydrodynamics with with conformal equation of state, give similar results within few percent \cite{Kurkela:2018vqr,Liyanage:2022nua,Ambrus:2022koq}. It would be very difficult to identify
a sensitive experimental probe of an early,  short, non-equilibrium stage in the dynamics of  heavy-ion collisions based on the flow in the transverse direction only.

\begin{figure}
 \begin{center}
   \includegraphics[scale=0.5]{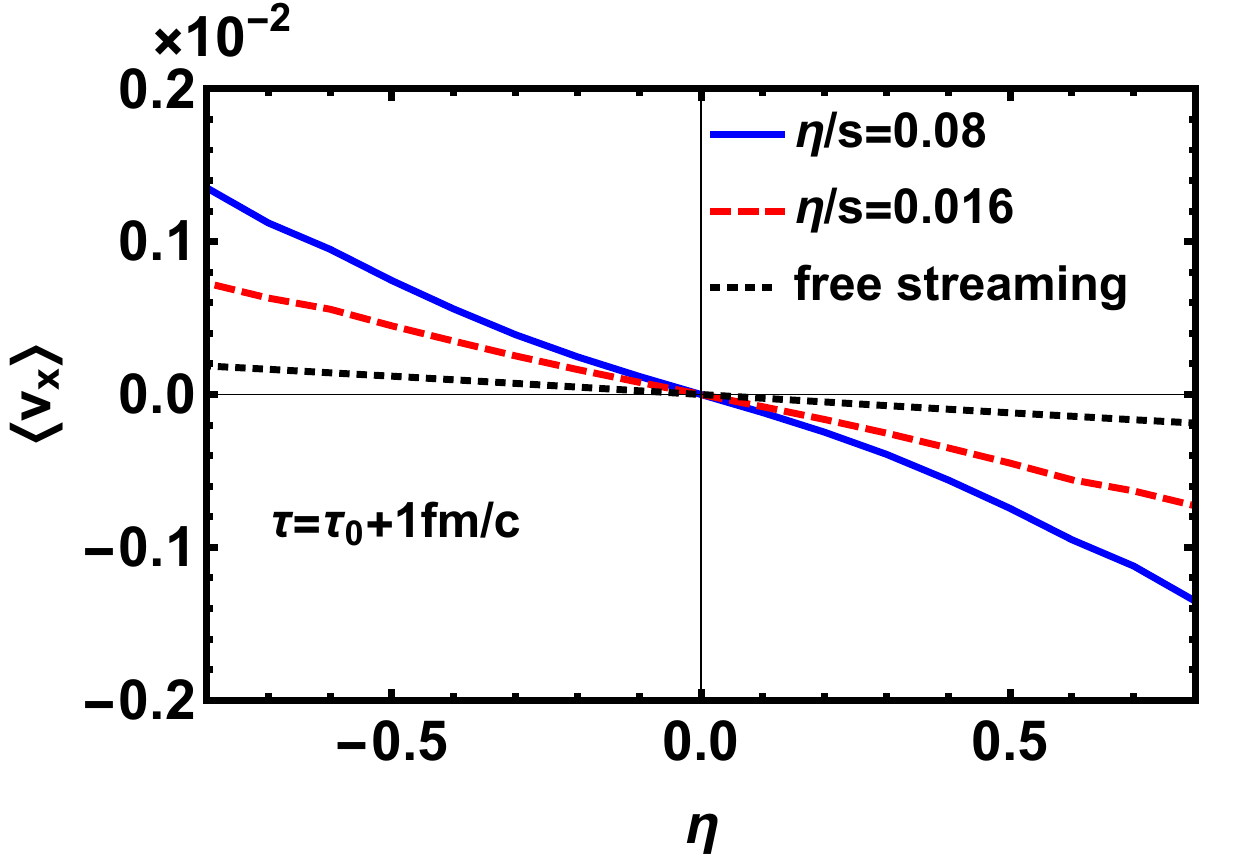}
   \vskip 0mm
 \caption{ The average velocity $\langle v_x\rangle $ as a function of space-time rapidity  (same symbols as in Fig. \ref{fig:plptcenter}).  }
 \label{fig:vxeta}
 \end{center}
\end{figure}

\begin{figure}
 \begin{center}
   \includegraphics[scale=0.5]{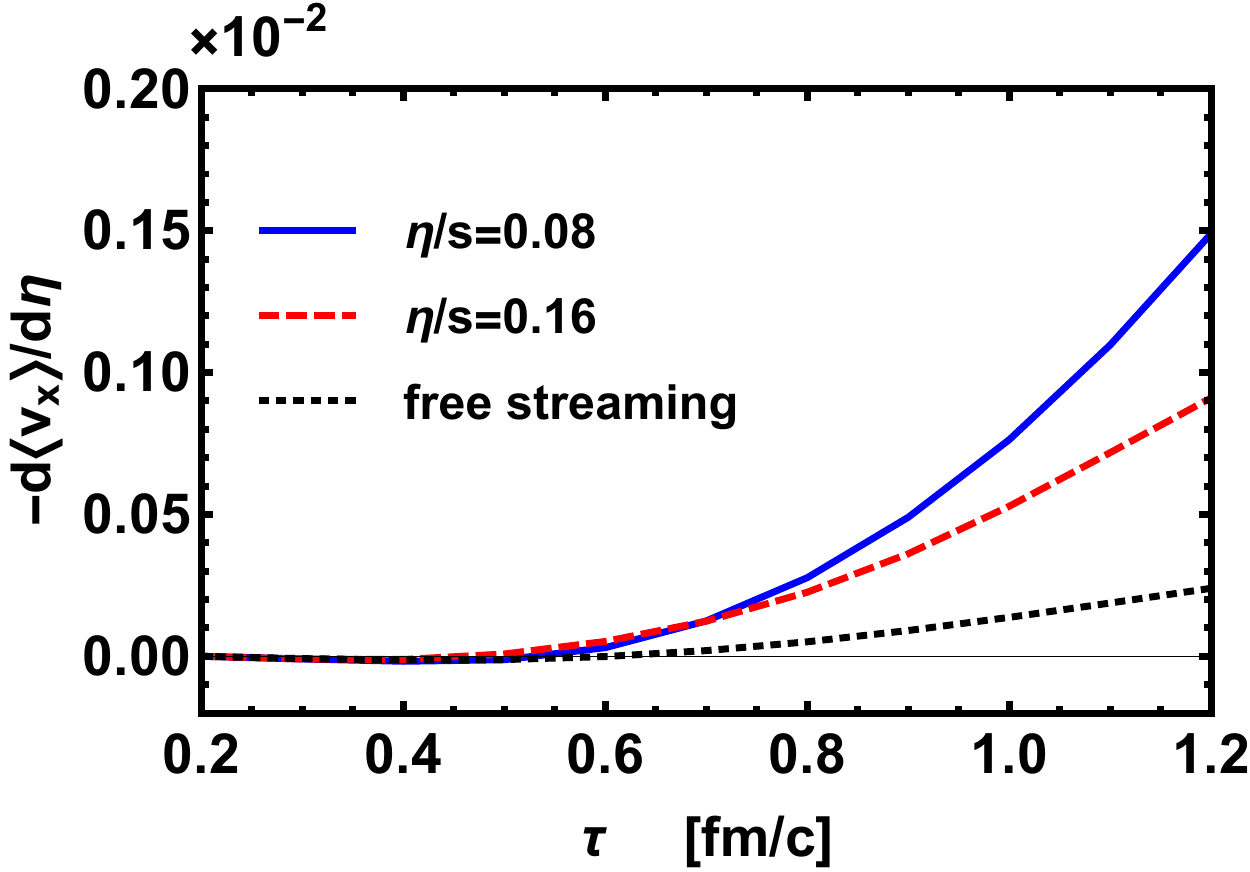}
   \vskip 0mm
 \caption{ The slope of the average velocity $\frac{- d\langle v_x\rangle}{d \eta} $ as  a function of the evolution time $\tau$ (same symbols as in Fig. \ref{fig:plptcenter}).  }
 \label{fig:dvxeta}
 \end{center}
\end{figure}

The simultaneous longitudinal and transverse expansion of the tilted
source initial conditions (Fig. \ref{fig:idens}) can lead to the formation
of a net asymmetric flow in the transverse direction at nonzero rapidities
\cite{Bozek:2010bi}. This asymmetric flow can be estimated by averaging the net
velocity $\langle v_x \rangle$
in the transverse  direction $x$  (the transverse
direction in the reaction plane). The longitudinal
flow remains close to the  Bjorken flow $Y \simeq \eta$ in the evolution.
The average
$\langle v_x \rangle$ calculated at fixed $\eta$ and $\tau$ is an estimate
of the net flow velocity of matter  with rapidity $Y$ at the proper time
$\tau$.

After the early kinetic evolution a net rapidity flow velocity
$\langle v_x \rangle $ in formed (Fig. \ref{fig:vxeta}).
The net flow velocity average  is very different in the three cases,
depending on the collision rate in the kinetic evolution. The largest flow asymmetry is generated in the evolution with
 the short relaxation time ($\eta/s=0.08$) and the smallest flow asymmetry for the free-streaming dynamics. The dependence
of the flow
velocity $\langle v_x \rangle $ on space-time rapidity
$\eta$ is approximately linear.
In Fig. \ref{fig:dvxeta} is shown the slope
$ - \frac{d\langle v_x \rangle}{d \eta}$ of the dependence
on space-time rapidity  as a function of the evolution time. It is
interesting to note that the  slope increases
in time in a nonlinear way, unlike the average transverse
flow (Fig. \ref{fig:vt}). The net flow generated in the early kinetic
evolution is sensitive to the collision rate  and to the degree of
equilibration between the longitudinal and transverse pressures in the system.

\section{Conclusions}

The kinetic evolution equations are studied in a system without boost
invariance. The chosen geometry, with dependence on
one transverse and one longitudinal direction, allows for a nontrivial
interplay of the longitudinal and transverse expansions. The kinetic equations
are solved for massless particles in the isotropization time approximation,
taking into account the full dependence on angles in the particle momentum
distribution. The initial conditions in the numerical solution break the
forward-backward symmetry in the longitudinal direction and allow for the
formation
of a rapidity dependent, asymmetric  flow in the transverse direction.

The transverse flow profile generated in the expansion is very similar,
independently on the chosen collision rate (relaxation time). The resulting
average transverse flow is almost the same, independently on  the value of the
equilibration rate, even though
the  pressure asymmetry  in the evolution is very different.
A quantity sensitive to the degree of non-equilibrium in the expansion
is given by the net flow in the transverse direction at non-zero space-time
rapidities. The calculation shows that the rapidity-dependent asymmetry of the
flow in
the transverse  direction changes significantly between the three cases
studied. The largest flow asymmetry is found  for a small relaxation time and the smallest asymmetry
for the free-streaming expansion.

The study  indicates that observables sensitive to the longitudinal
pressure should used, in order to test the early non-equilibrium
dynamics in heavy-ion collisions. One example of such a
quantity is the rapidity dependent
directed flow with respect to the reaction plane. In terms of modeling of the
collision dynamics, it requires the solution of the kinetic equation
in a non-boost invariant geometry.  The implementation of an effective solution
of the kinetic equation in $3+1$-dimensions would require the use of a
relativistic cascade model instead of the solution of the partial
differential kinetic
equation.  A realistic study should include a viscous hydrodynamics stage and
hadronization in order to make predictions for experimental observables. Finally, note that the kinetic equations solved  in this paper can
be  used to study the dynamics of the initial flow going beyond the
Bjorken scaling flow. The relaxation of a general initial  flow to the Bjorken
one ($Y\simeq \eta$) may influence the effective stopping of
matter in the dynamics
and indirectly also the 
collective flow observables.

\section*{Acknowledgments}

This research is partly supported by
the National Science Centre Grant No. 2018/29/B/ST2/00244.

\bibliography{../hydr}

\end{document}